\documentstyle[11pt,moriond,epsfig]{article}
\def\Journal#1#2#3#4{{#1} {\bf #2}, #3 (#4)}

\def\NPB{{\em Nucl. Phys.} B}
\def\PLB{{\em Phys. Lett.}  B}

\def\PRD{{\em Phys. Rev.} D}
\def\ZPC{{\em Z. Phys.} C}

\def\be{\begin{equation}}
\def\ee{\end{equation}}
\def\bea{\begin{eqnarray}}
\def\eea{\end{eqnarray}}

\begin{document}
\title{ANALYSE DES DONNEES COMBINEES SLAC-BCDMS-NMC A GRAND  ${\bf x}$:
${\bf \alpha_s}$ ET "HIGH TWISTS".}

\author{ S.I. ALEKHIN}
\address{l'Institute de la Physique des Hautes Energies, Protvino, 
142284, Russie}

\maketitle\abstracts
  {Nous avons fait une analyse en QCD, a NLO l'approximation, de la combinaison
des data de cross-sections profondes non-elastiques SLAC-BCDMS-NMC  avec 
determination de $\alpha_s$. Nous montrons que la 
valeur obtenue pour  $\alpha_s$ depend de la procedure 
statistique du traitement des erreurs systematiques.
L'ajustement des donnees avec la prise en compte complete  
des correlations  point-a-point donne une valeur
$\alpha_s(M_Z)=0.1183\pm0.0021(exp.)\pm0.0013(echelle de ren.)$
compatible avec les mesures de LEP et la moyenne mondiale.
Nous avons extrait d'une facon independante des modeles les 
contributions "high twist" aux fonctions de structures  
$F_L$ et $F_2$: la contribution de "twist-4" a  $F_L$
est en qualitatif accord avec les predictions du modele 
"renormalon" infrarouge; celle de " twist-6" a $F_L$ 
tend faiblement vers les valeurs negatives et celle de  
"twist-6" a $F_2$ vers  les valeurs positives, quoiqu'elles
soient toutes les deux compatibles avec zero compte tenu des 
erreurs.}

\newpage
\vspace*{4cm}
\title{COMBINED ANALYSIS OF SLAC-BCDMS-NMC DATA \\ AT HIGH 
${\bf x}$: ${\bf \alpha_s}$
AND HIGH TWISTS}

\author{ S.I. ALEKHIN}

\address{Institute for High Energy Physics, Protvino, 142284, Russia}

\maketitle\abstracts
{We perform a NLO QCD analysis of the nonsinglet part of 
the combined SLAC-BCDMS-NMC data on inclusive deep inelastic 
cross section with the extraction of 
$\alpha_s$. We show that the value of $\alpha_s$ obtained 
in the analysis is sensitive to the statistical inference 
procedures dealing with systematic errors on the data.
The fit with the complete account of point-to-point correlations 
of the data gives the value of 
$\alpha_s(M_Z)=0.1183\pm0.0021(exp.)\pm0.0013(ren.scale)$ that is
compatible with the  LEP measurements and the world average. 
Model independent x-shape of 
high twist contributions to the structure functions $F_L$ and $F_2$
is extracted. Twist-4 contribution to $F_L$ is
found to be in qualitative agreement with the predictions of
infrared renormalon model. Twist-6 contribution to $F_L$
exhibit weak trend to negative values, and twist-6 contribution to 
$F_2$ - to positive 
values, although both are compatible with zero within errors.}

Data on deep inelastic scattering (DIS) of charged leptons off fixed targets
\cite {SLAC,BCDMS,NMC}
are an unique source of information about nucleon structure and 
value of strong coupling constant $\alpha_s$. These data are obtained 
using high integral luminosity samples and their 
typical statistical errors are $O(1\%)$. 
However the experimental uncertainties are dominated by systematic errors
that typically are 2-3 times as statistical ones. Systematic errors 
are more difficult to be accounted since they are correlated
and an estimators which involve correlated data are complicated. 
This is the 
reason, why in many analysis systematic errors are accounted using 
simplified approach and/or partially (or even completely) ignored. 
The aim of our present
study is to perform QCD analysis 
of the data from Refs.\cite {SLAC,BCDMS,NMC} with a particular 
attention to thorough account of point-to-point correlations due to
systematic errors. 
A reliable estimate of $\alpha_s$ implies the study of 
possible influence of high twist (HT) contribution
to the scaling violation. In our analysis we perform simultaneous and 
model independent extraction of HT contribution to $F_L$ and $F_2$
and study their correlations with the $\alpha_s$ value. 

To allow for extraction of $F_L$
we analyzed the data on cross sections separated by the beam energies
instead of merged data on $F_2$. 
We imposed a cut of $x\ge0.3$
to prevent additional uncertainties due to a poorly 
known gluon distribution. This cut leaves 
data that to a good approximation can be described by 
the pure nonsinglet structure functions, which essentially
reduces the number of fitted parameters.
The cut $x\le0.75$, omitting the region where
the binding effects in deuterium are large
was also imposed in the analysis. 
The total number of the data points (NDP) left after the cut is 1243;
the $Q^2$ range of the data is $1~$GeV$^2<Q^2<230~$GeV$^2$;
the total number of independent systematic errors is 47. 

The QCD input leading-twist (LT)
structure functions of 
the proton and neutron were
parametrized at the starting 
value of $Q_0^2=9~$GeV$^2$ as follows: \footnote{We checked 
that extra polynomial-type
 factors do not improve 
the quality of the fits.}
\begin{displaymath}  
F^{p,n}_2(x,Q_0)=A_{p,n}x^{a_{p,n}}(1-x)^{b_{p,n}},
\end{displaymath}  
and then were evolved over the region of $Q^2$ occupied 
by the data in NLO QCD approximation 
in the $\overline{MS}$ factorization scheme \cite{MS}
with the help of the code used earlier.\cite{PDF} 
The final formula for the structure functions used in the fit, with account 
of the twist-4 contribution was chosen in an additive form
\begin{displaymath}  
F_{2,L}^{(p,d),HT}(x,Q)=F_{2,L}^{(p,d),TMC}(x,Q)+H_{2,L}^{(p,d)}(x)
\frac{1~{\rm GeV}^2}{Q^2},
\label{eqn:f2l_ht}
\end{displaymath}  
where $F_{2,L}^{(p,d),TMC}(x,Q)$ are given by NLO QCD 
with the account of target-mass correction.\cite{TMC}
The functions $H_{2,L}^{(p,d)}(x)$ were parametrized in a model-independent 
way: their values at $x=0.3,0.4,0,5,0.6,0.7,0.8$ were fitted;
between these points the functions were linearly interpolated.
The data on differential cross sections were fitted using the formula
\begin{displaymath}
\frac{d^2\sigma}{dxdy}=\frac{4\pi\alpha^2(s-M^2)}{Q^4}
\biggl[\biggl(1-y-\frac{(Mxy)^2}{Q^2}\biggr)F_2^{HT}+
\biggl(1-2\frac{m_l^2}{Q^2}\biggr)
\frac{y^2}{2}\biggl(F_2^{HT}-F_L^{HT}\biggr)\biggl],
\end{displaymath}
where $s$ is total c.m.s. energy, $m_l$ is scattered lepton mass and 
$y$ is lepton scattering variable.

To take into account the point-to-point correlations of the data points
we minimized a functional 
\begin{displaymath}
\chi^2=\sum_{K,i,j}
(f_i/\xi_K-y_i)E_{ij}(f_j/\xi_K-y_j), 
\end{displaymath}
where $K$ runs through the 
data subsets obtained by separation of all analyzed data on 
experiments and targets, and $i$ through data points within these subsets.
The matrix $E_{ij}$ is the inverse of the covariance matrix
$C_{ij}=\delta_{ij}\sigma_i\sigma_j+f_if_j(\vec{s}_i^K \cdot \vec{s}_j^K)$,
and each vector $\vec s_i^K$ includes all independent systematic errors
for the $K$-th data subset. The other notations are:
$y_i$ = the measurements; $\sigma_i$ = the statistical errors; 
$f_i$ = the theoretical model 
prediction depending on the fitted parameters. The normalization 
factors $\xi$ were fitted for old SLAC experiments and fixed at 1
for BCDMS, NMC, and SLAC-E-140.

\begin{figure}
\centerline{\psfig{file=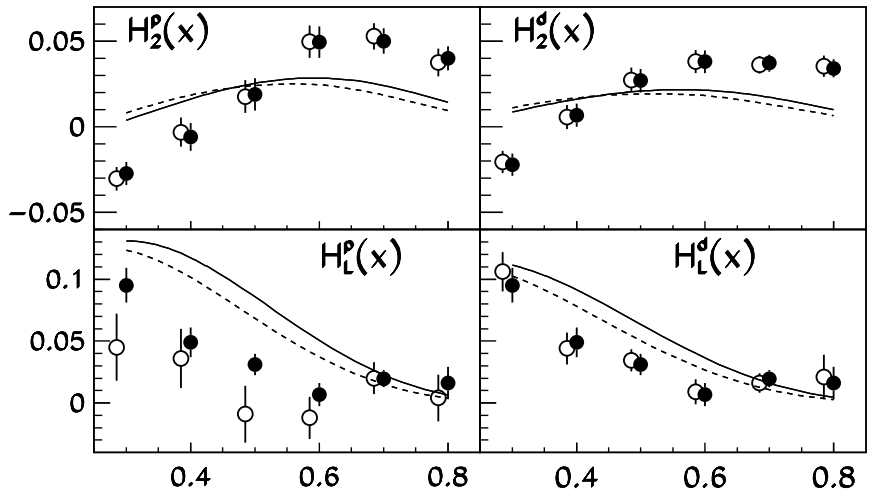,height=6.5cm}}
\caption{The fitted values of $H_2(x)$ and $H_L(x)$ 
(full circles -- the fit with constraint 
$H_L^D(x)=H_L^P(x)$, empty circles -- the unconstrainted fit).
The curves are predictions of IRR model
for $Q^2=2~$GeV$^2, n_f=3$ (full lines) 
and $Q^2=9~$GeV$^2, n_f=4$ (dashed lines).}
\label{fig:ht2l}
\end{figure}

The results of this fit are given in Fig.~\ref{fig:ht2l}. 
In view of large errors of $H^P_L(x)$ we imposed 
the constraint $H^P_L(x)=H^D_L(x)$ in the final fit.
The statistical quality of the fit is acceptable:
$\chi^2/$NDP$=1255/1243$.
The value of strong coupling constant obtained is 
$\alpha_s(M_Z)=0.1170\pm0.0021$(stat+syst)
that correspond to 
$\Lambda^{(3)}_{\overline{MS}}=337\pm29$(stat+syst)~MeV or 
$\Lambda^{(4)}_{\overline{MS}}=301\pm30$(stat+syst)~MeV.
 Pure statistical error of $\alpha_s(M_Z)$ is 0.0011, which 
gives only small contribution to the total experimental error.
The average bias of the  
the fitted function against data, calculated as
$\Biggl\langle(f-y)/\sqrt{\sigma^2+f (\vec{s})^2}\Biggl\rangle$
is 0.07 that is within its 
possible statistical fluctuation. The correlation 
of $\alpha_s$ with the HT contribution to $F_2$ is very large
(typical values of correlation coefficients is about --0.9).
This means that separation of logarithmic and power effects in 
the analysis of scaling violation, which is based on  
the SLAC-BCDMS-NMC data without $Q^2$ cut, 
is unstable under various assumptions. 
In particular, the complete account of 
point-to-point correlations of the data leads to a 
shift of the $\alpha_s$ value by about 3 standard 
deviations from the results obtained using a
simplified statistical inference procedure.\cite{VIR}
Since the HT contribution and the value of $\alpha_s$ are
strongly anticorrelated, the increase of $\alpha_s$
is accompanied by a decrease of HT.\footnote{This 
effect was also recently observed in the analysis,\cite{YB} 
where $\alpha_s(M_Z)$ was fixed at 0.120.}
The total effect on the HT magnitude is about a factor of 3/4, 
as compared with the results of Ref.\cite{VIR}.
At the same time $H_L$ is almost uncorrelated with $\alpha_s$, i.e.
its value is less model dependent.
The predictions of infrared renormalon (IRR) model \cite{IRR}
are also given in Fig.~\ref{fig:ht2l}. The normalization 
factor $A'_2$ was chosen in the universal form: 
$A'_2=-\frac{2C_F}{\beta_0}\bigl[\Lambda_R\bigr]^2e^{-C}$
where $C_F=4/3$, $C=-5/3$, $\beta_0=11-2/3n_f$,
$\Lambda_R=\Lambda^{(3)}_{\overline{MS}}=337~$MeV, as obtained
in our analysis. 
One can see that the model qualitatively describes the data on $H_L(x)$ 
and there is evident discrepancy with the data on $H_2(x)$.  

\begin{figure}
\centerline{\psfig{file=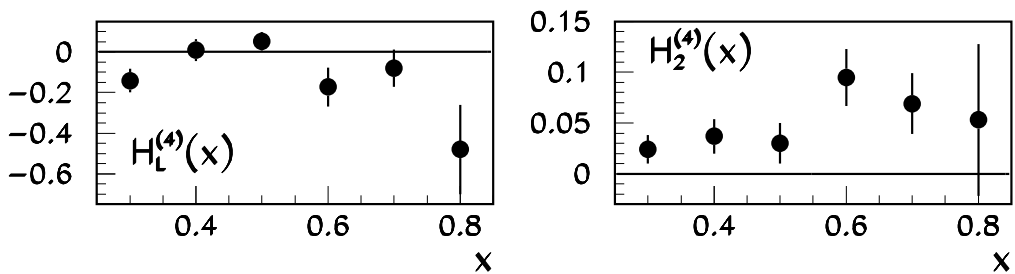,height=3.5cm}}
\caption{The dependence of twist-6 contributions to $F_L$ 
and $F_2$ on $x$.}
\label{fig:ht4}
\end{figure}

We checked how much the analyzed data are sensitive to the 
twist-6 contribution to $F_L$ and $F_2$. For this purpose we 
added to $F_2^{HT}$ or $F_L^{HT}$ the terms
$H_{L,2}^{(4)}(x)\frac{1~{\rm GeV}^4}{Q^4}$, where functions $H_{L,2}^{(4)}(x)$
were the same for proton and deuterium, and parametrized similarly 
to $H_{L,2}(x)$. The fitted values of $H_{L,2}^{(4)}(x)$ 
are given in Fig.~\ref{fig:ht4}.
One can observe the trend to the negative values at highest $x$ for 
$H_{L}^{(4)}(x)$ and trend to the positive values for $H_{2}^{(4)}(x)$.
However the statistical 
significance of the deviation of $H_{L,2}^{(4)}(x)$ from zero is not
very large. In addition, the correlation of $H_2^{(4)}(x)$ 
with $\alpha_s$ and $H_2(x)$ is extremely strong.
Summarizing these observations, we can conclude that the 
observed deviation of $H_L^{(4)}(x)$ and $H_2^{(4)}(x)$ off zero
can be considered as qualitative only.  

\begin{figure}[t]
\centerline{\psfig{figure=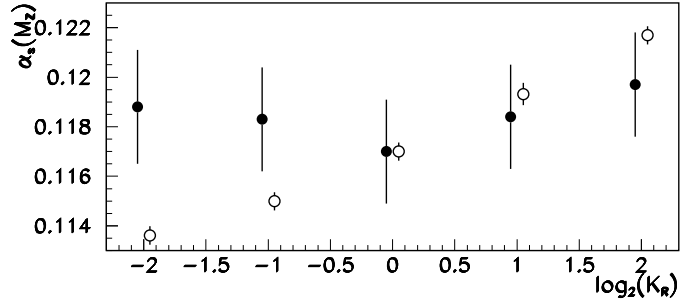,height=4cm}}
\caption{The dependence of $\alpha_s(M_Z)$ on the
renormalization scale. Open circles correspond to the fits with
the HT contributions fixed, full circles - to the fits with the HT 
contributions released. 
\label{fig:rena}}
\end{figure}

\begin{figure}
\centerline{\psfig{figure=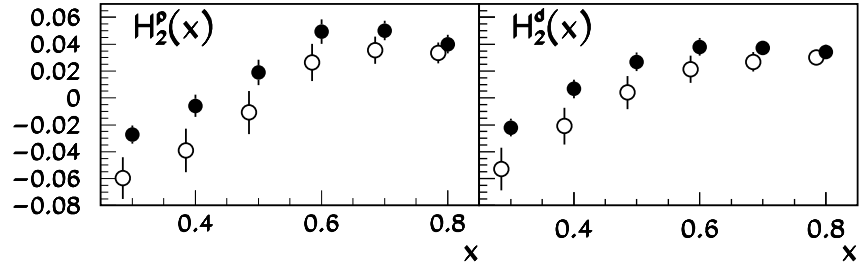,height=3.5cm}}
\caption{The values of HT contributions for $K_R=0.25$ (open circles)
and for $K_R=1$ (full circles).
\label{fig:renht}}
\end{figure}

Any renormalization group
analysis, which is  based on the finite number of perturbative series
terms is sensitive to the choice of renormalization scale.
This dependence can be used for the rough estimate of 
the higher orders terms effect that are not accounted for in the  
analysis; if one includes all terms this dependence should vanish.  
For estimate of renormalization scale uncertainty in DIS analysis
this scale usually is being chosen as $K_RQ^2$ and the value 
of $K_R$ is varying in the range of 0.1--4.\cite{RENSCALE} 
We studied the dependence of $\alpha_s$
on the value of $K_R$ in two cases: for the HT contributions released
in the fit and for the HT contributions fixed at the 
values obtained in the fit with $K_R=1$.
The results are given in Fig.~\ref{fig:rena}. One can see that 
the dependence of $\alpha_s$ on 
the renormalization scale is different in the two cases.  
The reason of this difference is that the $H_2^{p,d}(x)$ are strongly 
correlated with $\alpha_s$; when one changes 
$K_R$, the changes in the QCD evolution kernel
can be absorbed into the additional power-like contribution. 
This effect, illustrated in Fig.~\ref{fig:renht}, can be 
considered as indirect indication that the fitted values of 
$H_2^{p,d}(x)$ can include not only genuine power corrections,
but also effectively account for higher orders QCD terms. 
Any way the the value of $\alpha_s$ is quite stable against
renormalization scale variation, if HT are released.
At the same time the $\alpha_s$ errors are significantly larger
for this case, which is also a consequence 
of the large correlation of $\alpha_s$ with HT, i.e.
one can say that the uncertainty in the $\alpha_s$ value 
due to the renormalization scale choice is partially 
hidden in the total experimental error.
The spread of the $\alpha_s$ value in the fits with 
$K_R=$0.25--4 and HT released is 0.0026. With the account 
of this spread we obtain
$\alpha_s(M_Z)=0.1183\pm0.0021$(exp.)$\pm0.0013$(ren.scale).


\end{document}